\documentclass[preprint,showpacs,preprintnumbers,amsmath,amssymb]{revtex4}%
\usepackage{graphicx}
\usepackage{dcolumn}
\usepackage{bm}
\usepackage{amsmath}
\usepackage{epstopdf}
\usepackage{amsfonts}
\usepackage{amssymb}%
\providecommand{\U}[1]{\protect\rule{.1in}{.1in}}
\newcommand{\be}{\begin{equation}}
\newcommand{\en}{\end{equation}}
\newcommand{\bea}{\begin{eqnarray}}
\newcommand{\ena}{\end{eqnarray}}
\begin{document}
\title{Natural Inflation on the brane with a TeV-scale gravity: Parameter constraints after Planck 2015}
\author{Nelson Videla}
\email{nelson.videla@ing.uchile.cl}
\affiliation{Departamento de F\'{\i}sica, FCFM, Universidad de Chile, Blanco Encalada 2008, Santiago, Chile}
\author{Grigorios Panotopoulos}
\email{grigorios.panotopoulos@tecnico.ulisboa.pt}
\affiliation{CENTRA, Instituto Superior T\'ecnico, Universidade de Lisboa,
Av. Rovisco Pa\'is 1, Lisboa, Portugal}
\date{\today}

\begin{abstract}
In the present work we have studied Natural Inflation in the framework
of the Randall-Sundrum II brane model (RS-II) in the light of the latest Planck
results. Adopting the Randall-Sundrum fine-tuning, the model is
characterized by 3 parameters in total, namely the 5-dimensional Planck
mass $M_5$ and the two mass scales of the inflaton potential $f$ and $\Lambda$.
We show in the $n_s-r$ plane the theoretical predictions of the model together with the allowed
contour plots, and we conclude that the model is viable. By using the Planck results
only it is possible to determine the two mass scales of the inflaton potential in terms
of $M_5$, which remains undetermined. However, there are several good theoretical reasons to consider
a higher-dimensional Planck mass of the order of $10 TeV$, which is compatible with primordial nucleosynthesis.
If we insist on considering a $M_5$ of this order of magnitude all parameters are known and a sub-Planckian excursion
of the inflaton scalar field is achieved.
\end{abstract}

\maketitle



\section{Introduction}

The inflationary universe has become one of the central paradigms in modern cosmology. This is due to the fact that
many long-standing problems of the Big Bang model, such as the horizon, flatness, homogeneity and monopole problems, find a natural
explanation in the framework of the inflationary universe \cite{R1,R106,R103,R104,R105,Linde:1983gd}. However, the essential
feature of inflation is that it generates a
mechanism to explain the Large-Scale Structure (LSS) of the universe \cite{R2,R202,R203,R204,R205}
and provides a causal interpretation of the origin of the anisotropies observed in the
Cosmic Microwave Background (CMB) radiation\cite{astro,astro2,astro202,Hinshaw:2012aka,Ade:2013zuv,Ade:2013uln}, since
primordial density perturbations may be produced from quantum fluctuations
during the inflationary era.

Despite its success, a theory of inflation is still missing, all we have is a large collection of inflationary models (see e.g.\cite{Lyth:1998xn}, and for a classification of inflationary models see\cite{Kinney:2003uw}). The inflaton
potential cannot be derived from a fundamental theory in
a unique way. Moreover, the potential must be specially
designed to be flat, and this is the fine-tuning problem of inflation.
Natural Inflation (NI) with a pseudo-Nambu Goldstone boson (pNGB) as
the inflaton\cite{Freese:1990rb} is provided in certain particle
physics model\cite{Adams:1992bn}. The scalar potential, which is flat due to shift
symmetries, has the form
\begin{equation}
V(\phi)=\Lambda^4\left[1-\cos\left(\frac{\phi}{f}\right)\right].\label{Vnat}
\end{equation}
and it is characterized by two mass scales $f$ and $\Lambda$ with $f \gg
\Lambda$. It is assumed that a global symmetry is spontaneously broken at
some scale $f$, with a soft explicit symmetry breaking at a lower scale $\Lambda$.
Natural inflation has been already studied in standard cosmology based on
Einstein's General Relativity\cite{Savage:2006tr}. Upon comparison to the current cosmological and astronomical observations, specially those related with the CMB
temperature anisotropies, it is possible to constrain several inflation models. Particularly, the constraints in the $n_s-r$ plane
give us the predictions of a number of representative inflationary
potentials. Recently, the Planck
collaboration has published new data of enhanced precision of
the CMB anisotropies \cite{Ade:2015lrj}. Here, the Planck full mission
data has improved the upper bound on the tensor-to-scalar ratio
$r_{0.002} < 0.11$($95\%$ CL) which is similar to obtained from  \cite{Ade:2013uln} , in which
$r < 0.12$ ($95\%$ CL). In particular, Natural Inflation is consistent with current data \cite{Ade:2013uln,Ade:2015lrj}
for trans-Planckian values of the symmetry breaking scale $f$, for which it may be expect the low-energy effective theory, on which (\ref{Vnat}) is based, to break down \cite{Banks:2003sx}. Regarding the last data of Planck, this model is consistent for $\log_{10}( f /M_p) > 0.84$, where $M_p$ is the reduced Planck mass $M_p\equiv 1/\sqrt{8\pi G_N}=2\times 10^8$ GeV. Nevertheless, several mechanisms which yield the potential (\ref{Vnat}) with super-Planckian values for the symmetry breaking scale $f$ consistent with a low energy description, have been proposed recently \cite{Silverstein:2008sg,McAllister:2008hb,Choi:2014rja,Harigaya:2014rga,Higaki:2014pja,Kappl:2014lra,Ben-Dayan:2014zsa,Long:2014dta,Gao:2014uha,Higaki:2014mwa}.

On the other hand, it would be challenging to study natural inflation in non-standard cosmologies.
Considering non-standard cosmologies is motivated by at least two facts, namely i) deviations from
the standard Friedmann equation arise in higher-dimensional theories of gravity, and ii)
there is no observational test of the Friedmann equation
before the primordial big-bang nucleosynthesis (BBN) epoch. A well-studied example of
a novel higher-dimensional theory is brane-world models. Brane models are
inspired from M/superstring theory and although
they are not yet derivable from the fundamental theory, at least they contain
the basic ingredients, such as extra dimensions, higher-dimensional objects
(branes), higher-curvature corrections to gravity (Gauss-Bonnet) etc. Since
superstring theory claims to give us a fundamental description of nature it is
important to study what kind of cosmology it predicts. Regarding the realization of
Natural Inflation in non-standard cosmologies, some
works have been put forward in the literature so far \cite{Felipe:2004wh, Furuuchi:2013ila, Neupane:2014vwa}, achieving sub-Planckian values for the symmetry breaking scale $f$, being consistent with the data available at that time.

The main goal of the present work is to study the realization of NI in the high-energy regime
of the RS-II brane model, in the light of the recent Planck
results. Later on we will show that our results are modified significantly
compared to\cite{Felipe:2004wh} using the Planck results. By comparing the
theoretical predictions of the model together with the allowed
contour plots, and we conclude that the model is viable. Using the latest Planck results only
the inflaton potential mass scales $f, \Lambda$ are given in terms of the five-dimensional Planck
mass, which remains unconstraint though. However, we insist on considering a higher-dimensional
Planck mass of the order of, say, $10 TeV$, since there are several good theoretical reasons for that.

We organize our work as follows: After this introduction, in the next section
we summarize the basics of the brane model as well as the dynamics of
inflation. In the third section we analyze natural inflation in the framework
of RS-II model and present our results, and in the last section we finish with
our conclusions. We choose units so that $c=\hbar=1$.

\section{Basics of Braneworld inflation}\label{branerew}

\subsection{Braneworld cosmology}

In the brane-world scenario the main idea is that our four-dimensional world and the standard model of particle physics are confined to live on a 3-dimensional brane, while gravity lives in the higher-dimensional bulk. Since the higher-dimensional Plank mass $M_D$ is the fundamental mass scale instead of the usual four-dimensional Planck mass $M_4$, the brane concept has been used to address the hierarchy problem of particle physics, initially in the simple framework of a flat (4+n) spacetime with 4 large dimensions and n small compact dimensions\cite{Antoniadis:1997zg}, and later refined by Randall and Sundrum\cite{Randall:1999ee, Randall:1999vf}. For an introduction to brane cosmology see e.g.\cite{Langlois:2002bb}. In the RS-II model \cite{Randall:1999vf}, the four-dimensional Einstein equations may be written as \cite{Shiromizu:1999wj}
\begin{equation}
^{(4)}G_{\mu \nu}=-\Lambda_4g_{\mu \nu}+\frac{8\pi}{M_4^2} \tau_{\mu \nu}+\left(\frac{8\pi}{M_5^3}\right)^2\pi_{\mu \nu}-E_{\mu \nu},\label{4DEEQ}
\end{equation}
where $\Lambda_4$ is the four-dimensional cosmological constant, $\tau_{\mu \nu}$ is the energy-momentum tensor of matter on the brane, $\pi_{\mu \nu}=(1/12) \tau \tau_{\mu \nu}+(1/8) g_{\mu \nu} \tau_{\alpha \beta} \tau^{\alpha \beta}-(1/4) \tau_{\mu \alpha} \tau_\nu^\alpha-(1/24) \tau^2 g_{\mu \nu}$, and $E_{\mu \nu}=C_{\beta \rho \sigma}^\alpha n_\alpha n^\rho g_\mu^\beta g_\nu^\sigma$ is the projection of the five-dimensional Weyl tensor $C_{\alpha \beta \rho \sigma}$ on the brane, where $n^\alpha$ is the unit vector normal to the brane. $E_{\mu \nu}$ and $\pi_{\mu \nu}$ encode the information about the bulk. The four-dimensional quantities can be computed in terms of the five-dimensional ones as follows\cite{Maartens:1999hf}
\begin{equation}
M_4=\sqrt{\frac{3}{4\pi}}\left(\frac{M_5^2}{\sqrt{\lambda}}\right)M_5\label{M5M4}
\end{equation}
\begin{equation}
\Lambda_4 = \frac{4 \pi}{M_5^3} \left ( \Lambda_5+\frac{4 \pi \lambda^2}{3M_5^3} \right) \end{equation}
The Friedmann equation for a flat FRW background is given by \cite{Binetruy:1999ut}
\begin{equation}
H^2=\frac{\Lambda_4}{3}+\frac{8\pi}{3M_4^2}\rho\left(1+\frac{\rho}{2\lambda}\right)+\frac{\mathcal{E}}{a^4}.\label{flatFE}
\end{equation}
where $\mathcal{E}$ is an integration constant coming from $E_{\mu \nu}$. The term $\frac{\mathcal{E}}{a^4}$ is known as the dark radiation, since it
decays in the same way as radiation. However, during inflation this term will be rapidly diluted, and we can neglect it. The five-dimensional Planck mass is constraint by the standard bib-bang nucleosynthesis to be $M_5 \geq 10 TeV$\cite{Cline:1999ts}.
In the following, we will take the four-dimensional cosmological constant $\Lambda_4$ to be zero, or in other words we adopt the RS fine tuning $\Lambda_5=-4 \pi \lambda^2/(3 M_5^3)$ so that model can explain the cosmic acceleration without cosmological constant, and neglecting the term $\frac{\mathcal{E}}{a^4}$  the Friedmann equation (\ref{flatFE}) becomes
\begin{equation}
H^2=\frac{8\pi}{3M_4^2}\rho\left(1+\frac{\rho}{2\lambda}\right),\label{finalflatFE}
\end{equation}
which becomes the basis of our study on brane-world inflation.

\subsection{Inflationary dynamics}

At low energies, i.e., when $\rho\ll \lambda$, inflation in the brane-world scenario behaves in exactly the same way as standard inflation. But at higher energies we would expect the dynamics of inflation to be changed.

We consider slow-roll inflation driven by a scalar field $\phi$, for which the energy density $\rho$ and the pressure $P$ are given by
$\rho=\frac{\dot{\phi}^2}{2}+V(\phi)$ and $P=\frac{\dot{\phi}^2}{2}-V(\phi)$, respectively, where $V(\phi)$ is the scalar potential. We assume that the scalar field is confined to the brane, so the four-dimensional Klein-Gordon equation still holds
\begin{equation}
\ddot{\phi}+3H\dot{\phi}+V^{\prime}=0,\label{KG}
\end{equation}
where prime indicates derivative with respect to $\phi$, and dot a derivative with respect to cosmic time. We can use the slow-roll
approximation to write  (\ref{finalflatFE}) and (\ref{KG}) as
\begin{equation}
H^2\simeq\frac{8\pi}{3M_4^2}V\left(1+\frac{V}{2\lambda}\right),\label{finalflatFESR}
\end{equation}
and
\begin{equation}
3H\dot{\phi}\simeq -V^{\prime}.\label{KGSR}
\end{equation}
In this way, using these two equations, it is possible to write the slow-roll parameters on the brane as\cite{Maartens:1999hf}
\begin{eqnarray}
\epsilon_V & \equiv & \frac{M_4^2}{16 \pi}\left(\frac{V^{\prime}}{V}\right)^2\frac{1+V/\lambda}{\left(1+V/2\lambda\right)^2},\label{ep}\\
\eta_V & \equiv & \frac{M_4^2}{8 \pi}\frac{V^{\prime \prime}}{V}\frac{1}{1+V/2\lambda}.\label{et}
\end{eqnarray}
Slow-roll inflation implies that $\epsilon_V\ll1$ and $\left|\eta_V\right|\ll1$, as in standard cosmology.
These reduce to standard slow-roll parameters at the the low-energy limit $V\ll \lambda$. On the other hand, in the high-energy limit, i.e., $V\gg \lambda$, these expressions become
\begin{eqnarray}
\epsilon_V &\simeq & \frac{ M_4^2\lambda}{4 \pi}\frac{V^{\prime\,2}}{V^3},\label{epHE}\\
\eta_V &\simeq & \frac{ M_4^2\lambda}{4 \pi}\frac{V^{\prime \prime}}{V^2}.\label{etHE}
\end{eqnarray}
The deviations from standard slow-roll inflation can be seen in the high-energy as both the parameters are suppressed by a factor $V/\lambda$.

The number of $e$-folds in the slow-roll approximation, using  (\ref{finalflatFE}) and (\ref{KG}), yields
\begin{equation}
N \simeq -\frac{8\pi}{M_4^2}\int_{\phi_{*}}^{\phi_{end}}\frac{V}{V^{\prime}}\left(1+\frac{V}{2\lambda}\right)\, d\phi,\label{Nfolds}
\end{equation}
where $\phi_{*}$ and $\phi_{end}$ are the values of the scalar field when the cosmological scales cross the Hubble-radius and at the end of inflation, respectively.
As it can be seen, the number of $e$-folds is increased due to an extra term
of $V/\lambda$. This implies a more amount of inflation, between these two values of the field, compared to standard
inflation.

\subsection{Perturbations}

In the following, we will give a review of cosmological perturbations in brane-world inflation. We consider the gauge invariant quantity $\zeta=-\psi-H\frac{\delta \rho}{\dot{\rho}}$. Here, $\zeta$ is defined on slices of uniform density and reduces to the curvature perturbation at super-horizon scales. A fundamental
feature of $\zeta$ is that it is nearly constant on super-horizon scales\cite{Riotto:2002yw}, and in fact this property does not depend on the gravitational field equations\cite{Wands:2000dp}. Therefore, for the spatially flat gauge, we have $\zeta=H\frac{\delta \phi}{\dot{\phi}}$, where $\left|\delta \phi\right|=H/2\pi$. In this way, using the slow-roll approximation, the spectra
of scalar perturbations is given by\cite{Maartens:1999hf}
\begin{equation}
\mathcal{P}_{\mathcal{R}}=\frac{H^2}{\dot{\phi}^2}\left(\frac{H}{2\pi}\right)^2 \simeq \frac{128\pi}{3 M_4^6}\frac{V^3}{V^{\prime 2}}\left(1+\frac{V}{2\lambda}\right)^3.\label{AS}
\end{equation}
On the other hand, the tensor perturbations are more involved since the gravitons can propagate in the bulk. The amplitude of tensor perturbations is given by\cite{Maartens:1999hf}
\begin{equation}
\mathcal{P}_{g}=\frac{64\pi}{M^2_4}\left(\frac{H}{2\pi}\right)^2F^2(x),\label{TS}
\end{equation}
where
\begin{eqnarray}
F(x) &=& \left[\sqrt{1+x^2}-x^2\ln\left(\frac{1}{x}+\sqrt{1+\frac{1}{x^2}}\,\right)\,\right]^2\nonumber\\
&=& \left[\sqrt{1+x^2}-x^2 \sinh^{-1}\left(\frac{1}{x}\right)\right]^{-1/2},\label{FT}
\end{eqnarray}
and $x$ is given by
\begin{equation}
x=HM_4\sqrt{\frac{3}{4\pi \lambda}}.\label{xFT}
\end{equation}

The expressions for the spectra are, as always, to be evaluated at Hubble radius crossing $k = aH$.
As expected, in the the low-energy limit the expressions for the scalar and tensor spectra become the same as those derived without considering the brane effects. However, in the high-energy limit, these expressions become
\begin{eqnarray}
\mathcal{P}_{\mathcal{R}} &\simeq & \frac{16\pi}{3 M_4^6 \lambda^3}\frac{V^6}{V^{\prime \,2}},\label{ASHE}\\
\mathcal{P}_{g}  &\simeq & \frac{32 V^3}{M_4^4 \lambda^2}.\label{ATHE}
\end{eqnarray}

The scale dependence of the scalar power spectra  is determined by the  scalar spectral index, which under the slow-roll approximation, obeys the usual relation
\begin{eqnarray}
n_s &=&1+\frac{d \ln \mathcal{P}_{\mathcal{R}}}{d \ln k}\nonumber\\
n_s &\simeq & 1-6\epsilon_V+2\eta_V. \label{ns}
\end{eqnarray}

The amplitude of tensor perturbations can be parameterized by the tensor-to-scalar ratio, defined to be\cite{Peiris:2003ff}
\begin{equation}
r\equiv \frac{\mathcal{P}_{g}}{\mathcal{P}_{\mathcal{R}}},\label{rr}
\end{equation}
which implies that in the low-energy limit this expression becomes $r\simeq 16 \epsilon_V$, where $\epsilon_V$ is
the standard slow-roll parameter, whereas in the high-energy limit we have that $r \simeq 24 \epsilon_V$, with $\epsilon_V$ corresponding to Eq.(\ref{epHE}).

As we have seen, at late times the brane-world cosmology is identical to the standard one. During the early universe, particularly during inflation, there may be changes to the perturbations predicted by the standard cosmology, if the energy density is sufficiently high compared with the brane tension. In the following, we will obtain the predictions for the natural inflationary model in the brane-world scenario in the high-energy limit, and to try to ascertain whether these predictions are compatible with current observational constraints.

\section{Natural inflation on the brane}\label{natbra}

\subsection{Dynamics of inflation}

Natural inflation in the Randall-Sundrum brane-world scenario is characterized by 3 parameters in total, namely the 5-dimensional Planck mass and the mass scales of the inflaton potential (\ref{Vnat}), $\Lambda$ and $f$, respectively. In the high-energy limit, the slow-roll parameters $\epsilon_V$ and $\eta_V$, using Eqs.(\ref{epHE}) and (\ref{etHE}) are given by
\begin{eqnarray}
\epsilon_V =\alpha\frac{(1+\cos(y))}{(1-\cos(y))^2},\label{epHEV}\\
\eta_V =\alpha \frac{\cos(y)}{(1-\cos(y))^2},\label{etHEV}
\end{eqnarray}
where $y \equiv \phi/f$ and $\alpha$ is a dimensionless parameter defined as
\begin{eqnarray}
\alpha \equiv \frac{M_4^2 \lambda}{4\pi f^2 \Lambda^4}.\label{alpha}
\end{eqnarray}

For this model,
the condition for the end of inflation is found to be $\epsilon(y_{end})=1$, leading to
\begin{equation}
\cos(y_{end})=\cos \left(\frac{\phi_{end}}{f}\right)=\frac{1}{2}\left(2+\alpha-\sqrt{\alpha}\sqrt{8+\alpha}\right).\label{yend}
\end{equation}

The number of inflationary $e$-folds that occur between the values of the scalar field when a given perturbation scale leaves the Hubble-radius and at the end of inflation, can be computed from (\ref{Vnat}) and the high-energy limit of (\ref{Nfolds}), yielding
\begin{equation}
N=\frac{1}{\alpha}\left[\cos(y_{*})-\cos{(y_{end})}-2\ln\left(\frac{1+\cos(y_{*})}{1+\cos(y_{end})}\right)\right],\label{efoldsV}
\end{equation}
with $\cos(y_{end})$ given by (\ref{yend}). Solving Eq.(\ref{efoldsV}) for $\cos(y_{*})$, after replacing  Eq.(\ref{yend}), we may obtain the value of the scalar field at the time of Hubble-radius crossing, given by
\begin{equation}
\cos(y_{*})=\cos\left( \frac{\phi_{*}}{f}\right)=-1-2\, W\left[z(N,\alpha)\right],\label{ycmb}
\end{equation}
where
\begin{equation}
z(N,\alpha)\equiv -\frac{\sqrt{e^{-1-\frac{\sqrt{\alpha}}{2}(2N\sqrt{\alpha}+\sqrt{\alpha}-\sqrt{\alpha+8})}\left(\alpha^2+8 \alpha+8-\alpha^{3/2}\sqrt{\alpha+8}-4\sqrt{\alpha}\sqrt{\alpha+8}\right)}}{2\sqrt{2 e}},\label{zN}
\end{equation}
and $W$ denotes the Lambert $W$ function \cite{Veberic:2012ax}.

\subsection{Cosmological perturbations}

Regarding the cosmological perturbations, the amplitude of scalar perturbations, using Eqs.(\ref{Vnat}) and (\ref{ASHE}), is found to be
\begin{equation}
\mathcal{P}_{\mathcal{R}}=\frac{1}{12 \pi^2 \alpha^3}\gamma^4\frac{(1-\cos(y))^5}{(1+\cos(y))},\label{ASV}
\end{equation}
where $\gamma \equiv \frac{\Lambda}{f}$ is the ratio between both mass scales of the inflaton potential (\ref{Vnat}), $\Lambda$ and $f$. The scalar spectral index, using Eqs.(\ref{ns}), (\ref{epHEV}), and (\ref{etHEV}), becomes
\begin{equation}
n_s=1-2\alpha\frac{(3+2\cos(y))}{(1-\cos(y))^2}.\label{nsV}
\end{equation}

Finally, the tensor-to-scalar ratio can be obtained from the high-energy limit of Eq.(\ref{rr}), yielding
\begin{equation}
r=24\alpha\frac{(1+\cos(y))}{(1-\cos(y))^2}.\label{rrV}
\end{equation}

After evaluating these inflationary observables at the value of the scalar field when a given perturbation scale leaves the Hubble-radius, given by (\ref{ycmb}), we may compare the theoretical predictions of our model with the observational data in order to obtain constraints on the parameters that characterize it.

The amplitude of the scalar perturbations, the scalar spectral index, and the tensor-to-scalar ratio, evaluated at the Hubble-radius crossing $k=aH$, become
\begin{eqnarray}
\label{AsN}
\mathcal{P}_{\mathcal{R}} &=& \frac{4}{3 \pi \alpha^3}\gamma^4\frac{(1+W\left[z(N,\alpha)\right])^5}{\left(-W\left[z(N,\alpha)\right]\right)},\\
\label{nsN}
n_s &=&  1-\frac{\alpha}{2}\frac{\left(-W\left[z(N,\alpha)\right]\right)}{(1+W\left[z(N,\alpha)\right])^2} ,\\
\label{rrN}
 r &=& 12\alpha\frac{\left(-W\left[z(N,\alpha)\right]\right)}{(1+W\left[z(N,\alpha)\right])^2}.
\end{eqnarray}

\begin{figure}[th]
{\hspace{-3
cm}\includegraphics[width=4.8 in,angle=0,clip=true]{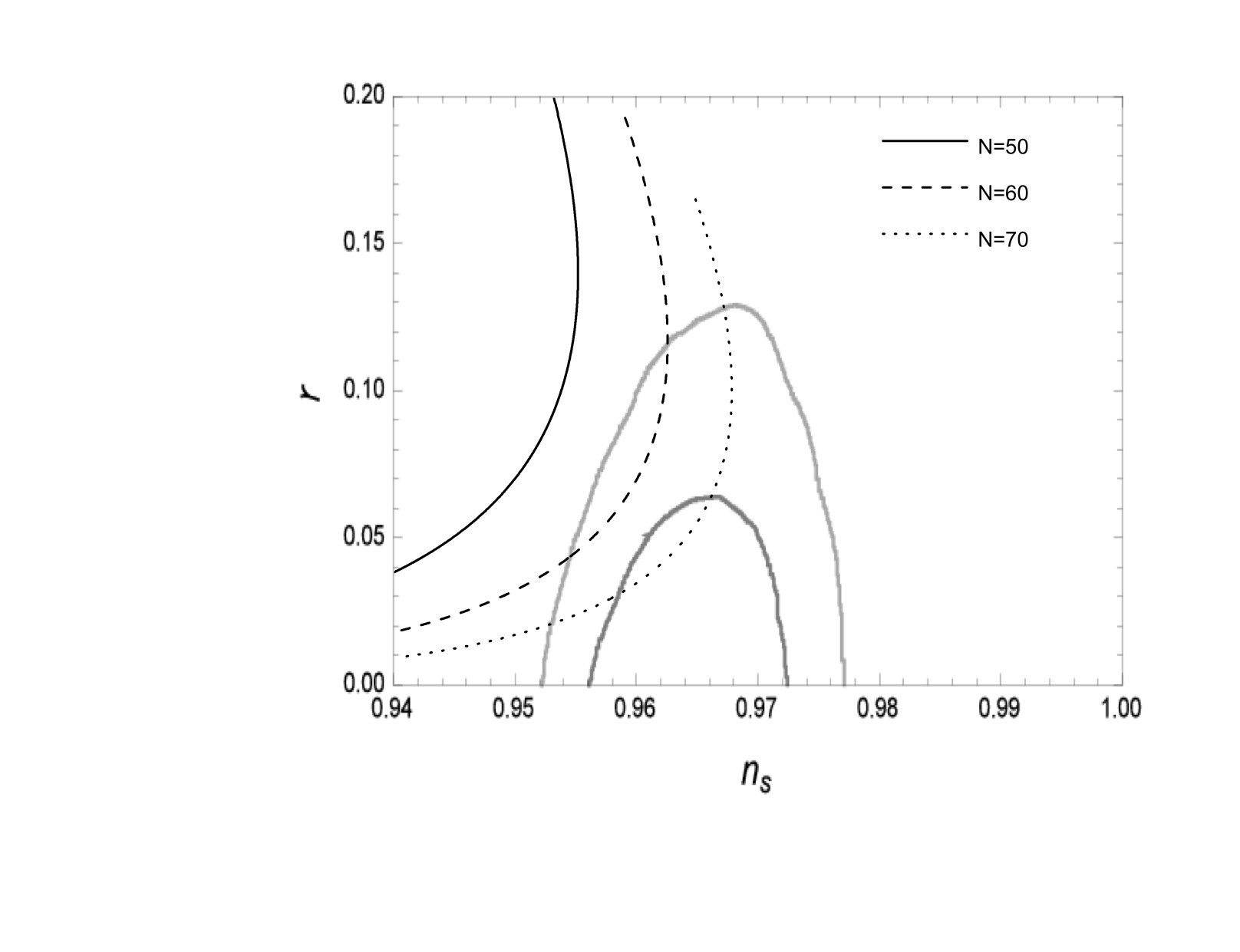}}
{\vspace{-1.5cm}\caption{Plot of the tensor-to-scalar ratio $r$ versus the scalar spectral index $n_s$ for natural inflation in the high-energy limit of braneworld scenario. Here, we have considered the
two-dimensional marginalized joint confidence contours for $(n_s
,r)$, at the $68\%$ and $95\%$ CL, from the latest Planck data \cite{Ade:2015lrj}. In this plot we
have used 3 different values for the number of $e$-folds $N$: the solid, dashed, and dotted lines correspond to $N=50,\,60$, and 70, respectively.
 \label{natrns}}}
\end{figure}

The trajectories in the $n_s$ - $r$ plane for the model studied here may be generated by plotting  Eqs.(\ref{nsN}) and (\ref{rrN}) parametrically. In particular, we have obtained three different curves by fixing the number of $e$-folds to $N=50,\,60$, and $70$, and plotting with respect to the parameter $\alpha$ in a wide range. The Fig.(\ref{natrns}) shows the plot of the tensor-to-scalar ratio $r$ versus the scalar spectral index $n_s$ for natural inflation in the high-energy limit of brane-world scenario. Here, we have considered the
two-dimensional marginalized joint confidence contours for $(n_s
,r)$, at the $68\%$ and $95\%$ CL, from the latest Planck data \cite{Ade:2015lrj}. The corresponding allowed range for the dimensionless parameter $\alpha$ defined by Eq.(\ref{alpha}), for each $r(n_s)$ curve, may be inferred by finding the points when the trajectory enters and exits the $95\%$ CL region from Planck. We can determinate numerically from Eqs.(\ref{nsN}) and (\ref{rrN})  that, by fixing $N$, the tensor-to-scalar ratio decreases as the parameter $\alpha$ is increasing. On the other hand, after to reach a maximum value, the scalar spectral index begins to decrease as $\alpha$ is increasing. In this way, we may obtain a lower limit as well an upper limit for $\alpha$.
The trajectory for $N=50$ lies well outside of the joint $95\%$ CL region in the $n_s$ - $r$ plane, so it is ruled out by the data. For $N=60$, the trajectory lies inside the $95\%$ CL region, obtaining the constraint on $\alpha$ given by $3\text{.}1\times 10^{-2}\lesssim \alpha \lesssim 7\text{.}1\times 10^{-2}$. Finally, for $N=70$ the trajectory lies inside the the joint $95\%$ CL region as well the $68\%$ CL region, obtaining the constraint on $\alpha$ given by $4\text{.}8\times 10^{-3}\lesssim \alpha \lesssim 8\text{.}5\times 10^{-2}$. On the other hand, by combining the scalar power spectrum (\ref{AsN}), the constraints on $\alpha$ already obtained, and the observational value for amplitude of the scalar power spectrum $\mathcal{P}_{\mathcal{R}}\simeq 2 \times 10^{-9}$ \cite{Ade:2015lrj}, me may obtain the allowed range for the ratio $\gamma \equiv \Lambda/f$ for each value of $N$. For $N=60$, this constraint becomes $6\text{.}6\times 10^{-3}\lesssim \gamma \lesssim 7\text{.}4\times 10^{-3}$, and for $N=70$ we have that $4\text{.}3\times 10^{-3}\lesssim \gamma \lesssim 6\text{.}5\times 10^{-3}$. Table (\ref{T1}) summarizes the constraints obtained on $\alpha$ and $\gamma$ using the last data of Planck.

\begin{table}
\centering
\begin{tabular}{|c|c|c|}
\hline
$N$ & constraint on $\alpha$ & constraint on $\gamma$ \\
\hline
60 & $3\text{.}1\times 10^{-2}\lesssim \alpha \lesssim 7\text{.}1\times 10^{-2}$ & $6\text{.}6\times 10^{-3}\lesssim \gamma \lesssim 7\text{.}4\times 10^{-3}$ \\
\hline
70& $4\text{.}8\times 10^{-3}\lesssim \alpha \lesssim 8\text{.}5\times 10^{-2}$ & $4\text{.}3\times 10^{-3}\lesssim \gamma \lesssim 6\text{.}5\times 10^{-3}$ \\
\hline
\end{tabular}
\caption{Results for the constraints on the parameters $\alpha$ and
$\gamma$ for natural inflation in the high-energy of Randall-Sundrum brane model, using the last data of Planck.} \label{T1}
\end{table}
As we can see, using the latest Planck results we find that the $\gamma$ parameter must be one order of magnitude larger than that found in\cite{Felipe:2004wh}, while the ratio $f/M_5 \sim 10$ (see equation (\ref{ff}) below) must be one order of magnitude lower compared to that
obtained in\cite{Felipe:2004wh}.

Clearly, we were not able to obtain the allowed range for all three
parameters of the model, namely the 5-dimensional Planck mass $M_5$ and the mass scales of the inflaton potential  $\Lambda$ and $f$, only by considering the Planck data. We have found the allowed range for certain combinations of these parameter, $\alpha$ and $\gamma$, instead. However, using the definitions of $\alpha, \gamma$ as well as the formulas relating the four-dimensional quantities with the five-dimensional ones, we can express everything in terms of the fundamental Planck mass $M_5$ as follows
\begin{equation} \
f = \left ( \frac{3}{16 \pi^2 \alpha \gamma^4} \right )^{1/6} M_5 \label{ff}
\end{equation}
\begin{equation}
\Lambda = \gamma f=\gamma \left ( \frac{3}{16 \pi^2 \alpha \gamma^4} \right )^{1/6} M_5 \label{LL}
\end{equation}
However, a string scale/higher-dimensional Planck mass of the order of a few TeV is very attractive from the theoretical point of view for several reasons, as it addresses the hierarchy problem, provides an alternative to gauge coupling unification in D-brane constructions of the Standard Model\cite{Antoniadis:2000ena}, and the evaporation a la Hawking of TeV mini-black holes can be seen at the colliders\cite{Casanova:2005id}, and possibly explain anomalies related to cosmic ray observations\cite{Mironov:2003jw}. Therefore, we shall take it seriously and in the rest of this article we shall assume for $M_5$ a value of $10 TeV$. Then, by combining the constraints found earlier on $\alpha$ and $\gamma$, and this value for the five-dimensional Planck mass, Eqs.(\ref{ff}) and (\ref{LL}) give us the allowed range for the spontaneous symmetry breaking scale $f$ and the soft explicit symmetry breaking scale $\Lambda$, yielding
$211.4 \,\text{TeV} \lesssim f \lesssim 261.9 \, \text{TeV}$ and $1.5 \,\text{TeV}\lesssim \Lambda \lesssim 1.8 \, \text{TeV}$ for $N=60$, whereas for $N=70$, the allowed ranges become
$223.7 \,\text{TeV} \lesssim f \lesssim 475.6 \, \text{TeV}$ and $1.3 \,\text{TeV} \lesssim \Lambda \lesssim 2.3 \, \text{TeV}$, respectively. These results imply a hierarchy between the mass scales consistent with $f\gg \Lambda$. On the other hand, the constraints found on $\alpha$ and Eqs.(\ref{yend}) and (\ref{ycmb}) imply that, during inflation $\phi\sim f$, therefore natural inflation in the framework of  the high-energy regime of the RS-II brane model takes place at sub-Planckian values of the scalar field.

\section{Conclusions}\label{conclu}

To summarize, in this article we have studied natural inflation in the framework
of the Randall-Sundrum II brane model in the light of the recent Planck
results. Adopting the Randall-Sundrum fine-tuning, the brane model is
characterized only by the 5-dimensional Planck
mass, while the inflationary model is characterized by the two mass scales $f, \Lambda$
of the inflaton potential. We have used the COBE normalization as well as the allowed contour plots
in the $n_s-r$ plane. First, in the $n_s-r$ plane we show the theoretical predictions of the model
for three different values of e-folds $N=50, 60, 70$.
According to the plot the $N=50$ case is excluded, while for the $N=60, 70$ cases
the model is viable for a certain range of the $\alpha$ parameter (a combination of the three
parameters of the model $f, M_5, \Lambda$) defined in the text. After that, using the constraint
for the amplitude of scalar perturbations we determined the ratio $\Lambda/f$. We have expressed
the mass scales of the inflaton potential in terms of the five-dimensional Planck mass which remains
unconstraint using the Planck results only.
It is known, however, that there are good theoretical reasons, such us hierarchy problem, alternative
to gauge coupling unification, mini black hole evaporation etc., to believe that $M_5$ is at the TeV scale.
If we take it seriously and insist on a $M_5 ~ 10 TeV$ all parameters of the model are known.


\begin{acknowledgments}
We would like to thank G. Barenboim for helping us with the figures. G.P. was supported by Comisi\'on Nacional
de Ciencias y Tecnolog\'ia of Chile through
Anillo project ACT1122. N.V. was supported by Comisi\'on Nacional
de Ciencias y Tecnolog\'ia of Chile through FONDECYT Grant N$^{0}$
3150490.
\end{acknowledgments}


\end{document}